\documentclass[jecats, manuscript]{copernicus}

\usepackage{subcaption}
\usepackage{tabularx}
\usepackage{booktabs}
\usepackage{makecell}
\usepackage{appendix}
\usepackage[T1]{fontenc}
\usepackage[utf8]{inputenc}

\newcolumntype{Y}{>{\centering\arraybackslash}X}

\begin{document}
\nolinenumbers


\title{Efficacy of Scalable Airline-led Contrail Avoidance}


\Author[1][tsankar@google.com]{Tharun}{Sankar} 
\Author[2]{Thomas}{Dean}
\Author[2]{Tristan}{Abbott}
\Author[3]{Jill}{Blickstein}
\Author[4]{Alejandra}{Martín Frías}
\Author[3]{Mark}{Galyen}
\Author[3]{Rebecca}{Grenham}
\Author[2]{Paul}{Hodgson}
\Author[1]{Kevin}{McCloskey}
\Author[3]{Alan}{Pechman}
\Author[2]{Tyler}{Robarge}
\Author[1]{Dinesh}{Sanekommu}
\Author[1]{Aaron}{Sarna}
\Author[1]{Aaron}{Sonabend-W}
\Author[2,5]{Marc}{Stettler}
\Author[4]{Raimund}{Zopp}
\Author[1]{Scott}{Geraedts}

\affil[1]{Google Research}
\affil[2]{Contrails.org}
\affil[3]{American Airlines}
\affil[4]{Flightkeys}
\affil[5]{Imperial College London}





\runningtitle{Efficacy of Scalable Airline-led Contrail Avoidance}

\runningauthor{T. Sankar et al.}

\maketitle



\begin{abstract}
Contrails account for a large portion of aviation's contribution to anthropogenic climate change. Navigational contrail avoidance is a promising solution to mitigate the warming caused by contrails. Prior trials testing navigational contrail avoidance have relied on bespoke integrations of contrail forecasts into airline operations. Here, we use a randomized control trial to test the feasibility of dispatcher-led contrail avoidance integrated into standard flight planning operations using a workflow that scales to an airline's entire network. We validated the efficacy of this intervention using satellite imagery and an automated flight-contrail attribution algorithm. Using this system, we observed an 11.6\% reduction in contrail formation rate for the 1232 flights marked as eligible for contrail avoidance (intent-to-treat) relative to the flights in the control group ($p = 0.011$). In the 112 flights that flew contrail avoidance as planned (per-protocol flights), we observed a 62.0\% lower contrail formation rate relative to the flights in the control group ($p < 0.001$). No statistically significant difference in fuel usage was observed between the two groups.
\end{abstract}

\section{Introduction}

Aviation contributes to anthropogenic climate change through a combination of carbon dioxide (CO$_2$) emissions and non-CO$_2$ effects \citep{lee2021contribution}. Among these non-CO$_2$ effects, contrail cirrus --- clouds formed by the condensation of water vapor on soot particles in the exhaust of aircraft engines --- is estimated to be the largest single component of aviation's net effective radiative forcing (ERF) \citep{karcher2018formation}. Unlike CO$_2$, which persists in the atmosphere for centuries, contrails are short-lived climate forcers, lasting from minutes to hours. This transience presents a unique opportunity for rapid climate mitigation: by avoiding the specific atmospheric conditions necessary for the formation of persistent warming contrails, the climate impact of aviation could be significantly reduced on short timescales \citep{teoh2020mitigating}.

The formation of persistent contrails requires an aircraft to fly through regions of the atmosphere that are supersaturated with respect to ice (Ice Supersaturated Regions, or ISSRs) and sufficiently cold \citep{schumann1996conditions}. Recent studies suggest that a small minority of flights are responsible for the vast majority of contrail warming. Theoretically, minor adjustments to flight altitudes or routes to avoid these regions --- a strategy known as navigational contrail avoidance --- could eliminate a significant portion of this warming with minimal fuel penalties \citep{teoh2020mitigating,caldeira2021contrails,frias2024feasibility,dean2025impact}.

Despite the theoretical promise, operationalizing contrail avoidance faces significant challenges, some of which are as yet unknown due to limited tests in real operational situations. Navigational avoidance requires accurately forecasting ISSRs in the upper troposphere and lower stratosphere. Recent studies have shown that the majority of contrail warming found in reanalysis products can be avoided through intelligent flight planning using forecast products at the lead times used by flight planners \citep{frias2024feasibility,dean2025impact}. However, these studies are based on simulations and do not capture whether the suggested contrail avoidance routes are operationally feasible. Additionally, several studies have called into question how well numerical weather prediction (NWP) models and their reanalysis products match real-world observations \citep{agarwal2022reanalysis,gierens2020well}. To address this, recent efforts have leveraged machine learning (ML) approaches trained on satellite observations to improve prediction accuracy \citep{sonabend2024feasibility}.

Prior operational trials have demonstrated the feasibility of navigational avoidance. For instance, \citet{sonabend2024feasibility} reported on a series of test flights where pilots altered climb and descent profiles based on contrail forecasts. Additionally, \citet{sausen2023can} conducted a randomized trial in the Maastricht Upper Area Control airspace involving tactical altitude adjustments by air traffic controllers. While these trials found that their interventions reduced the number of observed contrails, they relied on bespoke communication channels and manual interventions that do not scale to the high-volume, automated workflows of modern commercial aviation. Furthermore, previous studies sometimes relied on manual inspection of satellite imagery for validation, a process that is labor-intensive.

In this work, we present the results of a large-scale Randomized Control Trial (RCT) designed to test the feasibility and efficacy of scalable dispatcher-led contrail avoidance integrated directly into standard airline operations. Through a collaboration between American Airlines Inc. (AAL), Flightkeys GmbH, Contrails.org, and Google LLC, we integrated an ML-based contrail forecast into the Flightkeys GmbH platform (hereafter referred to as ``Flightkeys''), which AAL uses for creating flight plans in normal operations. This allowed for the generation of contrail-optimized flight plans across the airline's North Atlantic network without relying on labor-intensive planning methods, as the proposal of contrail-optimized flights plans was included in standard workflows.

The goal of this trial was to reduce observable contrail formation using the satellite-based validation system described in Sec.~\ref{sec:validation}, rather than relying on modeling results. Per-contrail estimates from existing contrail process models are not yet well-validated. Due to this, we selected a contrail forecast that focuses on contrail formation, with less emphasis on the specific warming impact of those contrails. We additionally included a climatological estimate of contrail warming described in Sec.~\ref{sec:forecast}. The climatological estimate causes our forecast to focus on contrails which are net warming and primarily during the evening or nighttime, since contrails during the daytime may have a net neutral or net cooling impact. We thereby minimized the risk that the trial increases contrail warming despite reducing persistent contrail formation.

We validated the efficacy of these avoidance maneuvers using an automated satellite verification system based on the work of \citet{sarna2025benchmarking}. This allows for an objective, blinded assessment of contrail formation at a scale previously unattainable. We compared the prevalence of observable contrails in flights that performed avoidance maneuvers against a control group of flights that did not consider contrails, providing statistically robust evidence of observable contrail reduction through scalable, integrated flight planning operations. We observed a 11.6\% reduction in contrail formation rate for the 1232 intent-to-treat flights relative to the control group, as well as a 62.0\% lower contrail formation rate for the 112 per-protocol flights relative to the control group.

\section{Methods}

\subsection{Contrail Forecast}
\label{sec:forecast}
The contrail forecast used during the flight planning process was produced by a ML system trained on contrail detections in the GOES-East geostationary satellite from the Advanced Baseline Imager (ABI) \citep{goodman2019goes}. The ML forecast model takes numerical weather forecast data from the European Center for Medium Range Weather Forecasts (ECMWF) as input and produces a probability of geostationary-observable contrail formation. This forecast system is similar to the one used in \citet{sonabend2024feasibility}.

For the training labels of the model, binary contrail labels were produced by a version of the system described in \citet{sarna2025benchmarking}. This system was modified slightly in this work to include a filtering based on altitude, where (flight, contrail) pairs are not considered for matching if the radiometrically estimated altitude of the contrail is more than 1615 m different from the advected altitude of the flight. More details are in Appendix~\ref{appendix:altitude}.

The probability of contrail formation was multiplied by a climatological estimate of contrail energy forcing obtained from the maps in \citet{platt2024effect}, resulting in a warming value in Joules per flight meter. The climatological map was generated by dividing the atmosphere into bins based on local time, latitude, and season, simulating all flights over a year in each bin using the Contrail Cirrus Prediction Model (CoCiP) \citep{schumann2012contrail}, and averaging the warming results. The result is a smoothly varying contrail impact map whose value is large for contrails expected to spend most of their lifetime at night when contrails are net warming, and small during the day when the warming impact is less certain. Incorporating this climatological map into our forecast allowed us to focus on the contrails likely to have a net warming effect, while still enabling the primary goal of reducing observable contrail formation. The forecast used for this trial was set to 0 for all locations East of $30^{\circ} W$ longitude since the GOES-East satellite was used for validation, and we did not want to perform contrail avoidance where it could not be validated. More details on the validation system are provided in Sec.~\ref{sec:validation}.

Contrail impacts are often considered in units of radiative forcing (RF), energy forcing (EF), effective radiative forcing (ERF), and CO$_2$-equivalent (CO$_2$e). During flight planning, contrail energy forcing was represented in units of tonnes of CO$_2$e using the central estimate ``ERF / RF'' ratio of 0.42 from \citet{lee2021contribution} and 100-year Absolute Global Warming
Potentials (AGWP100) \citep{IPCC_2021_WGI}, leading to a conversion factor of $3.36 \times 10^{12} J/tonne$.
Other methods of converting radiative forcing to CO$_2$e exist in the literature, and which method to use for different applications is a subject of active research \citep{Megill2024,borella2024}. In this work, one can think of CO$_2$e as an equivalent unit of energy forcing.

\subsection{Scope and Flight Selection}
\begin{figure}
 \centering
        \includegraphics[width=1\textwidth]{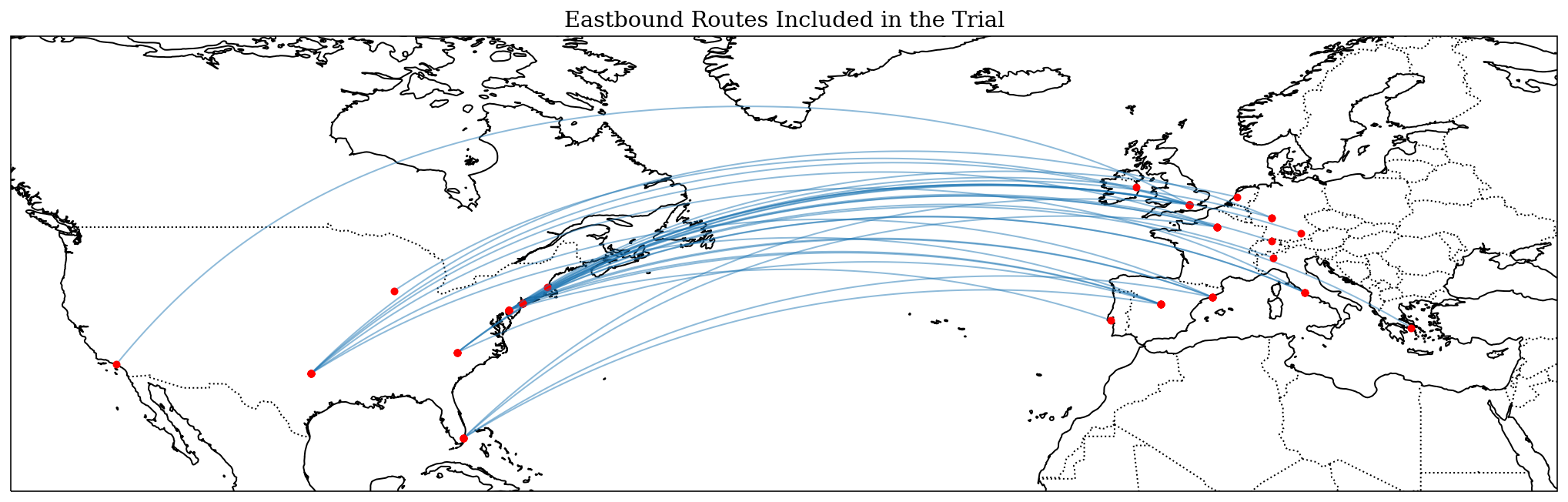}
 \caption{Eastbound flight routes considered for this trial. The displayed routes are great circle paths between each city pair and do not necessarily reflect the actual paths that were flown. A full list of city pairs included in the trial and their applicable time periods is provided in \autoref{tab:route_assignments}.}
\label{fig:routes_in_scope}
\end{figure}
For this trial, we considered all of AAL's eastbound flights from the United States to Europe, with certain flights removed from consideration due to operational constraints. All of the flights in the trial were part of AAL's standard schedule. The trial scope was explicitly defined as a set of origin-destination city pairs. Throughout the trial, 28 city pairs were considered, with certain city pairs added partway through the trial due to seasonal changes in AAL's flight schedule. \autoref{fig:routes_in_scope} shows a map of all city pairs considered for the trial. The eastbound flights to Europe primarily flew during the nighttime, meaning that any contrails produced by those flights were likely net warming. Therefore, this choice of scope also maximized the number of flights with high climatological contrail warming according to the forecast we used.

Every 2-3 weeks, each city pair was randomly assigned to the treatment or control group with a 50\% probability. The trial began with a ``ramp-up'' period, where city pairs were progressively added to the trial until week 5, when the trial included all candidate routes. A full list of city pairs and their applicable time periods is provided in \autoref{tab:route_assignments}. The Flightkeys system began by calculating the forecasted contrail EF for every flight in the treatment and control groups. This process began roughly 24 hours before the flight's departure time and was repeated multiple times before departure. If the accumulated contrail impact across the flight path was less than 10 tonnes (t) CO$_2$e, the flight was not considered for the trial. For flights in the control group that exceeded the 10 t threshold, operations proceeded as normal without considering contrails. Neither dispatchers nor pilots had any indication that a flight was included in the control group. These flights were later used for analysis to establish a baseline against which the treatment group was compared. For flights in the treatment group that exceeded the 10 t threshold, a contrail-optimized plan was produced using the Flightkeys optimizer. The 10 t threshold was chosen by examining simulated plans for a set of 95 pre-trial flights. The authors of this work manually chose an appropriate value that filtered out flights where the optimizer did not find a plan with significant contrail benefit.

The trial ran for 17 weeks beginning on January 15, 2025 and ending on May 13, 2025. The start and end dates were determined before the trial began and did not change depending on how many contrail avoidance flights were flown.

\subsection{Flightkeys Optimizer}
 The Flightkeys platform performs trajectory optimization using a gridded approach, leveraging established path-finding algorithms such as Dijkstra's and A* \citep{frias2024feasibility}. In this work, we refer to plans generated by the Flightkeys optimizer that do not consider contrails as ``non-avoidance'' plans.

For this study, forecasted contrail costs were included as an additional term in the cost function of the optimizer. The optimizer uses a marginal cost index to determine how heavily to weight contrails in the cost function. For this trial, we used a marginal cost value of \$50/tCO$_2$e. The Flightkeys platform coarse-grains the native $0.25^{\circ}$ contrail forecast resolution to $1^{\circ}$ by taking the average value. This resolution was chosen to balance between accuracy and compute resource usage. The platform receives hourly updates to the contrail forecast with a forecast range of 24 hours. In this work, we refer to plans generated by the Flightkeys optimizer that consider contrails as ``contrail-optimized'' plans.

\subsection{Operational Workflow}
Flights whose non-avoidance plan exceeded the 10 t threshold were re-optimized including the contrails cost term. The result of this procedure is a flight plan that minimizes a combination of contrail formation, fuel consumption, and other operational costs. Since the Flightkeys system is a cost-based system, the contrail-optimized plan may not avoid all forecasted contrails, if for example the extra fuel cost incurred by performing the contrail avoidance is too large.

In AAL's operations, dispatchers exercise joint operational control with pilots and are responsible for planning flights and working with the pilots to ensure safe operations for the duration of the flight. Once the contrail-optimized plan was generated, dispatchers were presented with a single contrail-optimized plan and one or more non-avoidance plans for consideration. Additionally, flights whose non-avoidance plan passed the 10 t threshold were indicated by a colored border in the Flightkeys user interface (UI), providing a visual cue for dispatchers to consider contrails for that flight. Dispatcher participation in the trial was voluntary, so the dispatcher for a given flight could decide to evaluate and release the contrail-optimized plan to the pilots or proceed as usual and evaluate and release a non-avoidance plan. If the dispatcher decided to release the contrail-optimized plan, the plan would be released with a tag indicating that contrails were considered for that flight and a remark informing the pilots that contrails were considered in the creation of the flight plan.

\subsection{Validation with Satellite Imagery} \label{sec:validation}
After the trial completed, we used a modified version of the ``contrail to flight'' attribution algorithm ``CoAtSaC'' (short for ``Contrail Attribution Sample Consensus'') \citep{sarna2025benchmarking}, with modifications described in \autoref{appendix:altitude}, to determine whether an observable contrail was created by each flight. This automated system also calculates the number of attributed observable contrail kilometers for each flight, which we used as the primary metric for comparing the treatment and control groups.

The system takes input images from the GOES-East geostationary satellite's ABI \citep{goodman2019goes}, and identifies contrails in those images using a machine learning computer vision algorithm~\citep{ng2023contrail}. The next step is to obtain Automatic Dependent Surveillance-Broadcast (ADS-B) flight path data. The data was a combination of terrestrial and satellite ADS-B data licensed from FlightAware LLC and Aireon LLC respectively. We process the ADS-B data using the process described in \citet{sarna2025benchmarking}. The system then attributes each flight to contrails by simulating the advection of the flight path using wind data from ECMWF ERA5 reanalysis \citep{hersbach2020era5}. It then compares the advected flight tracks to the detected contrails and determines that the flight matches an observed contrail if they are close enough, according to thresholds determined in \citet{sarna2025benchmarking}, with the difference that in this work we also considered whether the contrail and flights had similar altitudes (Appendix \ref{appendix:altitude}). The algorithm considers all spatiotemporally nearby flights for attribution, not only the flights included in the trial.

In order to keep the authors of this work blinded to the results, the satellite attribution was not performed until after the trial completed.

\subsection{Randomized Control Trial}
From the treatment group determined by random city pair assignment and 10 t warming threshold above, we considered three subgroups for analysis. These subgroups were defined before any members of the team were unblinded to the satellite results.
\begin{itemize}
    \item Level 1 (L1) - entire treatment group
    \item Level 2 (L2) - flights that were released as contrail-optimized plans
    \item Level 3 (L3) - flights that were released as contrail-optimized plans and flown as planned
\end{itemize}
The Level 1 group represents the full intent-to-treat group. The Level 2 and 3 groups are per-protocol groups, with differently strict definitions of protocol adherence. The Level 2 group is the subset of the L1 group where the dispatcher released the contrail-optimized plan to the flight crew. In order to determine which plan was released to the flight crew, we considered the last released plan before takeoff. The ratio of L2 to L1 flights (hereafter called the L2/L1 take rate, where ``take rate'' indicates the voluntary dispatcher engagement with contrail avoidance) was $15.4\%$. The Level 3 group is the subset of the L2 group where the flight adhered to the released flight plan. We considered a flight to have adhered to the contrail-optimized plan if the difference between the planned warming per flight km and the flown warming per flight km was less than 0.001 t CO$_2$e$/$km, where the warming was computed using the same forecast as described in \autoref{sec:forecast}. This threshold roughly corresponds to a 5 t CO$_2$e warming threshold for a 5000 km flight, which is the approximate length of many of the flights in the trial. This heuristic was determined before unblinding to the satellite results and was found to compare well with subjective human assessments of adherence performed by the authors of this work. These subjective assessments were performed by visualizing flight plans interpolated to the contrail forecast in the vertical, lateral, and time dimensions, and judging whether the flown ADS-B path matched the planned path. Examples of these visualizations are provided in Appendix~\ref{appendix:viz_examples}. This heuristic has the advantage of disregarding any non-adherence that was not near a contrail region. For example, if the flight climbed to its cruising altitude later than planned, but there were no contrail regions nearby, it could still be included in the L3 group assuming it performed the planned contrail avoidance maneuvers. The ratio of L3 to L1 flights (L3/L1 take rate) was $7.8\%$.

The reduction in contrail formation in the L1 group represents the average contrail reduction achieved over all the flights in the trial. However, it is strongly influenced by the relatively low take rate, which is in turn influenced by the design of this specific trial and may not be broadly applicable to other contrail avoidance efforts. The L2 group represents the contrail reduction achieved when dispatchers select and release a contrail avoidance flight plan. The L3 group represents the contrail reduction achieved when contrail avoidance routes were flown.

Each of these three treatment groups were evaluated against the control group using the observed per-flight contrail kilometers (normalized by flight distance) as the primary trial endpoint.

\subsection{Statistical Analysis} \label{sec:adjustment}

To compare contrail formation in the treatment and control groups, we computed the ratio of total observed contrail kilometers (computed as the length of the flight that is attributed to the observed contrails) normalized by total flight kilometers, referred to hereafter as the Observed Contrail Rate ($\lambda_C^{(k)}$):
\begin{equation}
\lambda_C^{(k)} = \frac{\sum_{i\in k} C_{i}}{\sum_{i\in k} L_{i}},
\label{lambdaC}
\end{equation}

\noindent where $i$ are the flights in either group $k$ (and $k$ is one of treatment or control), $C_{i}$ represents the contrail distance in kilometers, and $L_{i}$ represents the flight distance for flight $i$ in kilometers. When computing $L_i$, we used only the flight distance west of $30^\circ W$ longitude where the contrail forecast was available and where contrails could be observed by the GOES-East satellite.

To compare the fuel consumption between the treatment and control groups, we computed the Fuel Rate similarly to the Observed Contrail Rate (Eq.~\ref{lambdaC}):
\begin{equation}
\lambda_F^{(k)} = \frac{\sum_{i\in k} F_{i}}{\sum_{i\in k} L_{i}},
\label{lambdaF}
\end{equation}

where $F_i$ represents the fuel usage per flight and $L_i$ the total length of the flight (not restricted to west of $30^\circ W$). Fuel usage data is actual data provided by AAL. The quantity used for this analysis is the actual fuel used while the aircraft is in the air, derived by subtracting the fuel weight when the aircraft touches down from the fuel weight when the aircraft leaves the ground. This does not include the fuel used during the taxi phases before takeoff and after landing.

Due to operational constraints, city pairs were only randomized every 2-3 weeks. This caused an imbalance of aircraft type between treatment and control groups. For example, $74\%$ of the control group flights were Boeing 777s, whereas only $50\%$ of the treatment group flights were 777s. See Fig.~\ref{fig:aircraft_type_distribution} for the aircraft type distribution across groups.

The physical properties of a contrail are dependent on the combination of aircraft and engine type involved in contrail formation \citep{jessberger2013aircrafttype,unterstrasser2014}. However, the effect of these differences on geostationary-observable contrail formation is an area of active research \citep{driver2024factors,Gryspeerdt_2024}. Additionally, fuel consumption differs by aircraft type. Therefore, we treated aircraft type as a confounder for both contrail distance and fuel and adjusted the statistical tests accordingly. We report both marginal and adjusted results. To adjust the test statistic by aircraft type, we first computed the Observed Contrail and Fuel Rates per aircraft type. We then computed the weighted mean for treatment and control groups using the same weights, which are derived from the aircraft prevalence in the total study population. We computed the difference in both Observed Contrail and Fuel Rates: $\lambda^{(control)} - \lambda^{(treatment)}$ as our test statistics. Statistical significance was assessed using a non-parametric stratified permutation test with 10,000 permutations \citep{pesarin1993permutation,holt2023permutation}. To generate the null distribution, treatment labels were randomly shuffled among flights strictly within their respective aircraft type, preserving the aircraft-type confounding structure. A $p$-value was computed as the proportion of permuted statistics exceeding the observed reduction. For observed contrail formation, we used a one-sided $p$-value, and for fuel consumption we used a two-sided $p$-value.

To estimate the confidence interval, we performed a stratified bootstrap analysis with 10,000 bootstrap samples. We created new sample datasets by randomly drawing flights with replacement from the original treatment and control groups, doing so strictly within their respective aircraft strata to maintain the study's balance. For each iteration, we re-applied the aircraft prevalence weights and computed a new test statistic. The 95\% confidence interval was derived from the 2.5 and 97.5 percentiles these 10,000 simulated statistics.

\subsection{Counterfactual Analysis}  \label{sec:counterfactual_methods}
The L3 observable results in this work are the best estimate of how much reduction in contrail formation can be achieved with the methods used in this work when contrail-optimized routes are flown. The methods above compare the L2 and L3 groups to the control group. One drawback of the L2 and L3 analysis is that there might be some correlation between contrail formation and what causes a flight to be selected for the L2 and L3 groups. For example, in interviews conducted after the trial completed, dispatchers mentioned that the take rate depended on how busy they were. One factor that might make the dispatchers busy is more demanding weather conditions such as storms or turbulence, which could also be correlated with contrail formation. It may also be more difficult for the Flightkeys optimizer to offer a flyable contrail-optimized plan when a flight has to avoid large contrail-forming regions. In order to address these potential issues, we additionally performed a counterfactual analysis to confirm that the reduction in observed contrail km between the L2 and L3 groups is due to contrail avoidance. If the reduction factors from this counterfactual analysis were significantly smaller than the reduction factors in the observed results, we would have suspected that some amount of the observed reduction resulted from factors other than our intervention.

At many points in the flight planning process, the Flightkeys system generated both contrail-optimized and non-avoidance plans for a given flight. For flights where the contrail-optimized plan was released, we performed the additional counterfactual analysis to isolate the effect of our intervention without any additional confounders. To do this, we chose one of the non-avoidance plans as a counterfactual for the actual plan. In order to determine which non-avoidance plan to use as a counterfactual for the flight, we examined multiple choices of non-avoidance plan and compared them against the flown path for flights that did not perform contrail avoidance (specifically, flights from the treatment group that were not part of the L2 set, ``L1 not L2''). The choices we evaluated were the minimum cost plan and the most recent plan. To find the ``minimum cost'' plan, we examined all non-avoidance plans generated before the flight's departure and chose the plan with the minimum total cost as calculated by Flightkeys. The total cost is a combination of fuel and other costs such as delays and overflight charges \citep{frias2024feasibility}. The results of that analysis are included in Sec.~\ref{sec:counterfactual}.

We used the choice of non-avoidance plan from that analysis to compute the differences in expected contrail distance between the counterfactual plan and the flown path, where expected contrail distance was computed using the ML formation probability forecast described in Sec.~\ref{sec:forecast}. Since the expected contrail distance is a forecast product, we also validated that the forecast was well-calibrated against observations to justify its use in this analysis.

We additionally performed a counterfactual analysis for the fuel usage. Lacking simulated fuel data for the flown path, we compared the simulated fuel value for the last-generated contrail-optimized plan to the non-avoidance plan computed nearest to it in time, whether preceding or following.

\section{Results}

\subsection{Observable Contrails}
\begin{table}[ht]
\centering
\small 
\setlength{\tabcolsep}{2pt} 
\renewcommand{\arraystretch}{1.4}

\caption{Observable contrail formation results for the control and treatment groups. Reduction in contrail formation rate is calculated with respect to the contrail formation rate in the control group. The ``Adjusted" rate column incorporates the aircraft type adjustment described in Sec.~\ref{sec:adjustment}, whereas the ``Unadjusted" rate column does not. The values in brackets represent the 95\% confidence interval of the column quantity.}

\begin{tabularx}{\textwidth}{l Y Y Y Y Y}
\toprule

\textbf{Group} &
$N$ &
\makecell[b]{Contrail\\Formation Rate\\(Unadjusted)} & 
\makecell[b]{Reduction in Rate\\(Unadjusted)} &
\makecell[b]{Reduction in Rate\\(Adjusted)} &
\makecell[b]{$p$-value\\(Adjusted)} \\
\midrule

Control & $1172$ & \makecell[c]{$5.9\%$ \\ \scriptsize $[5.5\%, 6.4\%]$} & --- & --- & --- \\
\addlinespace
\makecell[l]{Treatment \\ (L1)} & $1232$ & \makecell[c]{$5.3\%$ \\ \scriptsize $[4.9\%, 5.6\%]$} & \makecell[c]{$\mathbf{11.6\%}$ \\ \scriptsize $[2.1\%, 20.1\%]$} & \makecell[c]{$\mathbf{11.6\%}$ \\ \scriptsize $[2.1\%, 20.3\%]$} & $\mathbf{0.011}$ \\
\addlinespace
\makecell[l]{Treatment \\ (L2)} & $190$ & \makecell[c]{$3.8\%$ \\ \scriptsize $[3.1\%, 4.5\%]$} & \makecell[c]{$\mathbf{36.4\%}$ \\ \scriptsize $[22.4\%, 49.0\%]$} & \makecell[c]{$\mathbf{29.5\%}$ \\ \scriptsize $[9.7\%, 46.0\%]$} & $\mathbf{0.002}$ \\
\addlinespace
\makecell[l]{Treatment \\ (L3)} & $112$ & \makecell[c]{$2.1\%$ \\ \scriptsize $[1.4\%, 2.8\%]$} & \makecell[c]{$\mathbf{64.5\%}$ \\ \scriptsize $[52.0\%, 75.7\%]$} & \makecell[c]{$\mathbf{62.0\%}$ \\ \scriptsize $[47.1\%, 75.3\%]$}& $\mathbf{<0.001}$ \\

\bottomrule
\end{tabularx}
\label{contrail_table}
\end{table}

Table \ref{contrail_table} shows the observed contrail formation rate for all the groups in our experiment. All subgroups of the treatment group showed a statistically significant reduction in observable contrail formation when compared against the control group. We rejected the null hypothesis of no change in observable contrail formation between the L1 treatment and control groups (one-sided $p=0.011$, stratified permutation test). We also rejected the null hypothesis of no change in observable contrail formation between the L2 treatment and control groups (one-sided $p=0.002$, stratified permutation test) and the L3 treatment and control groups (one-sided $p<0.001$, stratified permutation test). We see that the results are not substantially changed by the adjustment for aircraft type. Additional observed contrail values are included in Appendix~\ref{appendix:observed_values}.



\subsection{Contrail Warming Estimates}
 To estimate the warming reduction from our trial, we multiplied the observed contrail distance by the climatological warming estimates from \citet{platt2024effect}. 
The results of this analysis are shown in Table \ref{warming_table}. 
Once again we found significant reductions in all treatment subgroups. We rejected the null hypothesis of no change in climatological warming between the L1 treatment and control groups (one-sided $p=0.006$, stratified permutation test). We also rejected the null hypothesis of no change in climatological warming between the L2 treatment and control groups (one-sided $p<0.001$, stratified permutation test) L3 treatment and control groups (one-sided $p<0.001$, stratified permutation test). 


\begin{table}
\caption{Contrail warming estimates for the control and 3 treatment group levels. Reduction in contrail warming is calculated with respect to the contrail warming in the control group. The values in brackets represent the 95\% confidence interval of the column quantity.}
\centering
\small
\renewcommand{\arraystretch}{2}
\begin{tabularx}{\textwidth}{l Y Y Y}
\hline
\textbf{Group} &
\makecell{Climatological \\ Warming \\ (Joules / m, Unadjusted)} & 
\makecell{Reduction in \\ Climatological \\ Warming (Unadjusted)} &
\makecell{Reduction in \\ Climatological \\ Warming (Adjusted)} \\
\hline
Control & $9.20 \times 10^6$ & --- & --- \\
Treatment (L1) & $7.92 \times 10^6$ & \makecell[c]{$\mathbf{13.8\%}$ \\ \scriptsize $[2.6\%, 23.5\%]$} & \makecell[c]{$\mathbf{13.7\%}$ \\ \scriptsize $[3.1\%, 23.4\%]$}\\
Treatment (L2) & $5.78 \times 10^6$ & \makecell[c]{$\mathbf{37.1\%}$ \\ \scriptsize $[21.7\%, 50.4\%]$} & \makecell[c]{$\mathbf{35.5\%}$ \\ \scriptsize $[19.1\%, 50.0\%]$} \\
Treatment (L3) & $2.80 \times 10^6$ & \makecell[c]{$\mathbf{69.6\%}$ \\ \scriptsize $[56.5\%, 79.9\%]$} & \makecell[c]{$\mathbf{69.3\%}$ \\ \scriptsize $[56.1\%, 80.9\%]$} \\

\hline
\end{tabularx}
\label{warming_table}
\end{table}

\subsection{Fuel Usage} \label{sec:costs}
Contrail avoidance may incur additional costs, both monetary and climatological, primarily in fuel usage \citep{teoh2020mitigating,frias2024feasibility,dean2025impact}. This extra fuel is critical to quantify due to the climate impact from burning extra fuel. As previously noted, the Flightkeys optimizer used for generating contrail routes considers multiple costs, including fuel, contrails (with the specific price assignment for this trial), and others such as delays and overflight charges \citep{frias2024feasibility}.

\begin{table}[ht]
\caption{Fuel usage for the three treatment group levels. Differences in fuel usage are calculated with respect to the fuel usage in the control group. The ``Adjusted'' usage column incorporates the aircraft type adjustment described in \ref{sec:adjustment}, whereas the ``Unadjusted'' usage column does not. Negative differences indicate that the treatment group used less fuel than the control group, and positive differences indicate that the treatment group used more fuel.}
\centering
\small 
\renewcommand{\arraystretch}{1.4} 

\begin{tabularx}{\textwidth}{l c Y Y Y}
\toprule

\textbf{Group} &
$N$ &
\makecell[b]{Difference in\\Fuel Usage\\(Adjusted)} &
\makecell[b]{$p$-value\\(Adjusted)} &
\makecell[b]{Difference in\\Fuel Usage\\(Unadjusted)} \\
\midrule

Control & $1172$ & --- & --- & --- \\
\addlinespace
Treatment (L1) & $1232$ & \makecell[c]{$\mathbf{-0.55\%}$ \\ \scriptsize $[{-1.06\%}, {-0.03\%}]$} & \makecell[c]{$\mathbf{0.044}$} & \makecell[c]{$-7.2\%$ \\ \scriptsize $[{-8.5\%}, {-5.7\%}]$} \\
\addlinespace
Treatment (L2) & $190$ & \makecell[c]{$\mathbf{0.30\%}$ \\ \scriptsize $[{-0.65\%}, {1.28\%}]$} & \makecell[c]{$0.642$} & \makecell[c]{$-9.7\%$ \\ \scriptsize $[{-12.5\%}, {-6.9\%}]$} \\
\addlinespace
Treatment (L3) & $112$ & \makecell[c]{$\mathbf{0.02\%}$ \\ \scriptsize $[{-1.17\%}, {1.16\%}]$} & \makecell[c]{$0.974$} & \makecell[c]{$-10.9\%$ \\ \scriptsize $[{-14.7\%}, {-7.1\%}]$} \\

\bottomrule
\end{tabularx}
\label{fuel_table}
\end{table}

\begin{figure}[htbp]
     \centering
     \begin{subfigure}[t]{0.49\textwidth}
         \centering
         \includegraphics[width=\textwidth]{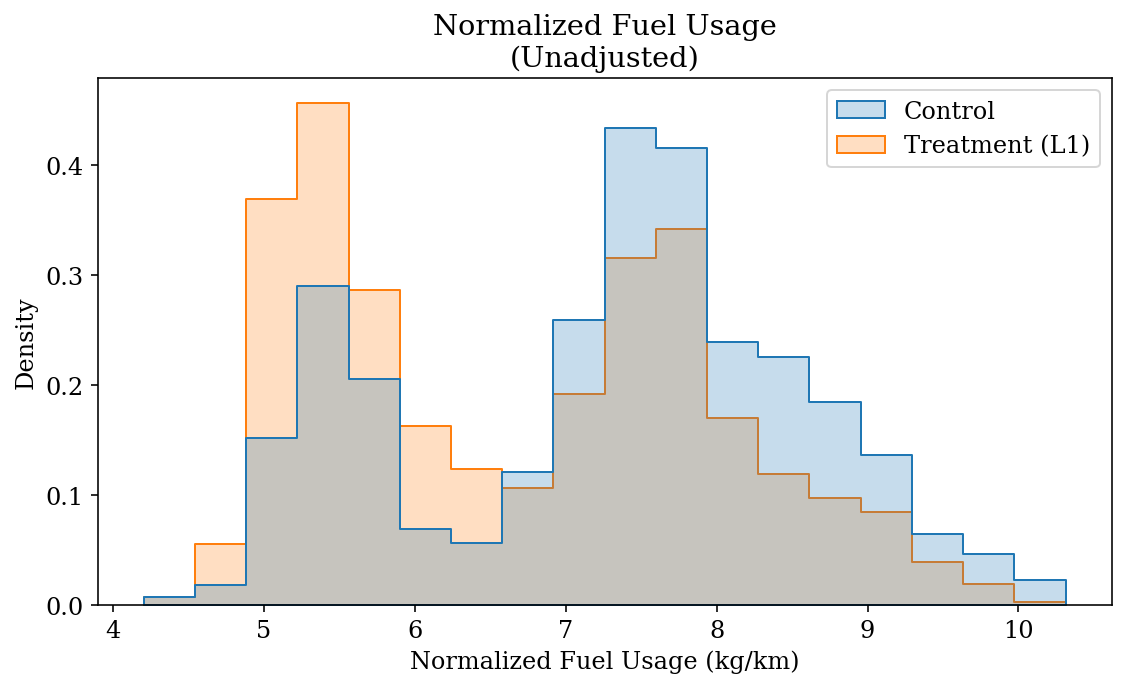}
         \caption{Normalized fuel usage before applying the aircraft type adjustment.}
         \label{fig:fuel_unadjusted}
     \end{subfigure}
     \hfill
     \begin{subfigure}[t]{0.49\textwidth}
         \centering
         \includegraphics[width=\textwidth]{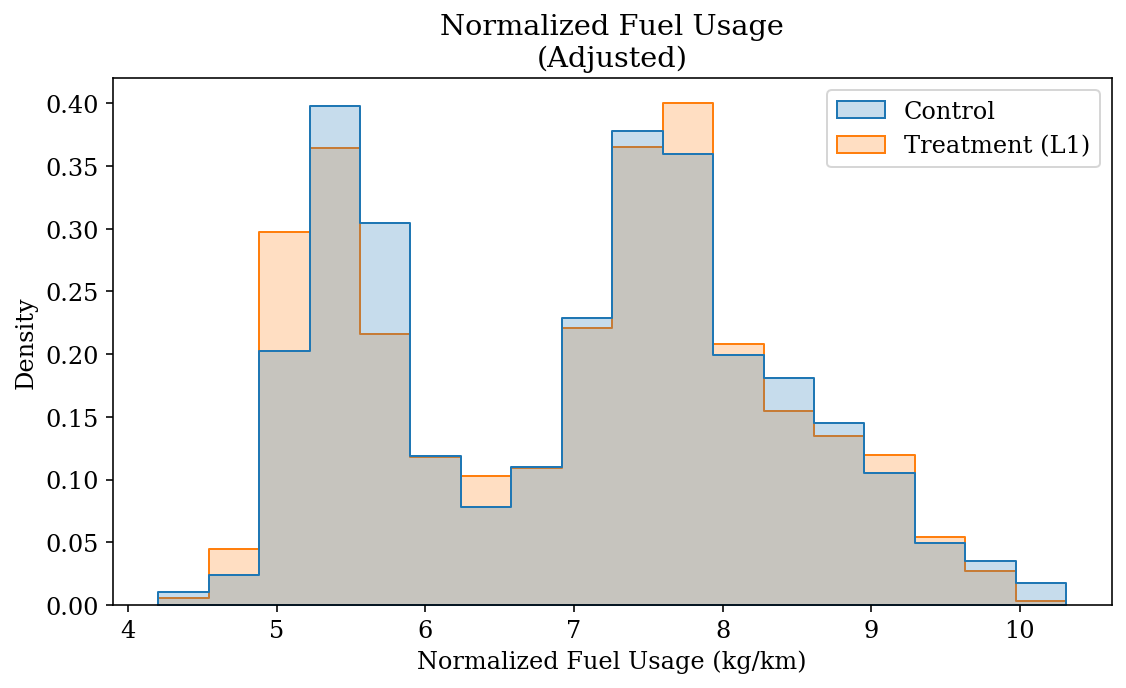}
         \caption{Normalized fuel usage after applying the aircraft type adjustment.}
         \label{fig:fuel_adjusted}
     \end{subfigure}
     
     \caption{Histogram of fuel/flight distance for each flight in the control and treatment groups. Two peaks correspond to the two different aircraft types in the trial: Boeing 777 and Boeing 787. In panel~\ref{fig:fuel_unadjusted}, the groups have different flight counts per peak; this imbalance is mostly removed in panel~\ref{fig:fuel_adjusted} after adjusting for aircraft type. Values in brackets represent the 95\% confidence interval.}
     \label{fig:fuel_aircraft_type_adjustment}
\end{figure}

Using actual fuel data provided by AAL, we compared the fuel usage between the control group and treatment subgroups. The results from this analysis are shown in \autoref{fuel_table}. The unadjusted numbers show a large reduction in fuel in the treatment group. The flights in the trial included two aircraft types: Boeing 777 and Boeing 787. \autoref{fig:fuel_aircraft_type_adjustment} visualizes the distribution of fuel per flight distance for each flight, demonstrating the aircraft type imbalance and effects of the adjustment described in \autoref{sec:adjustment}.

Even after the aircraft type adjustment, we were able to reject the null hypothesis of no change in fuel usage between the L1 treatment and control groups with relatively high confidence (two-sided $p=0.044$, stratified permutation test). However, the difference is small ($<1\%$) and the test is close to not being statistically significant at the $p=0.05$ level. The remaining difference may be due to other residual confounders such as different city pairs having slightly different average fuel usage, differences in weather, or differences in load factor. We are unable to reject the null hypothesis of no change in fuel usage between the L2 treatment and control groups (two-sided $p=0.642$, stratified permutation test) and the L3 treatment and control groups (two-sided $p=0.974$, stratified permutation test).



\subsection{Counterfactual Flightkeys-Generated Plans}  \label{sec:counterfactual}

\begin{figure}
 \centering
        \includegraphics[width=0.7\textwidth]{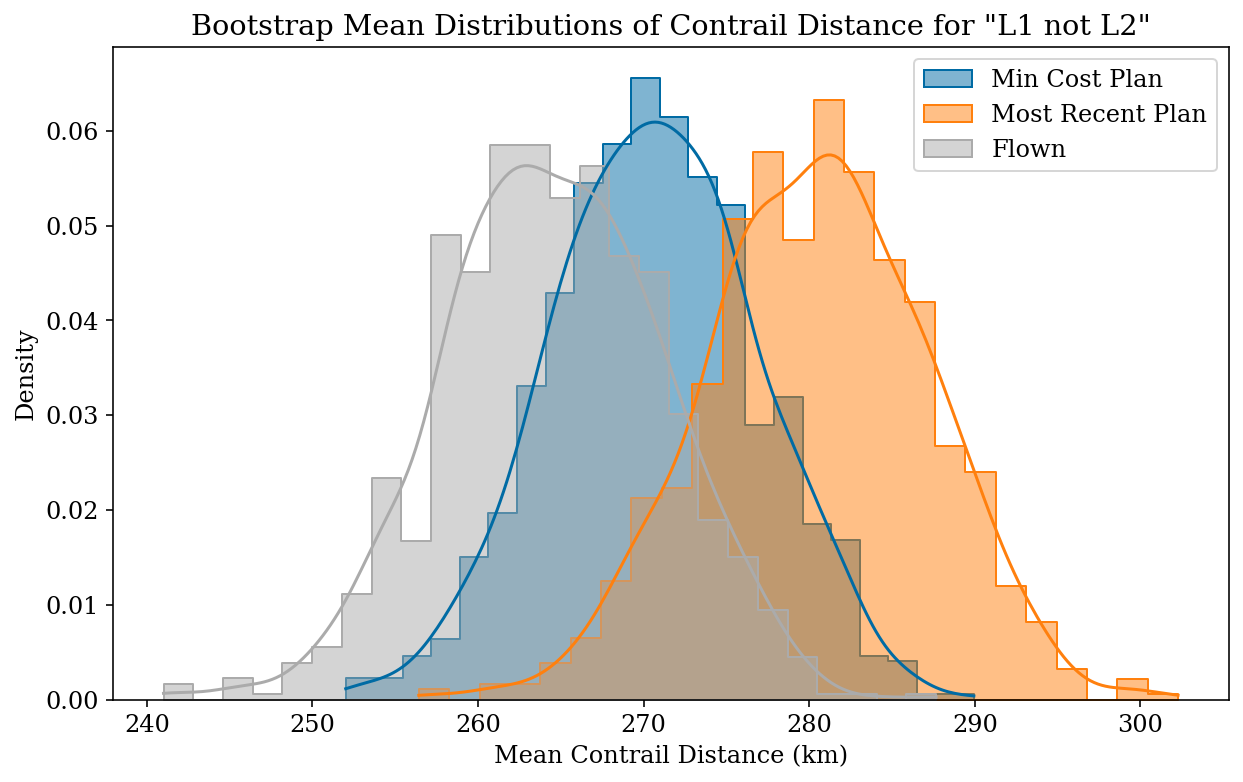}
 \caption{Bootstrap mean distributions of expected contrail distance for two potential choices of counterfactual plan (minimum cost plan and most recent plan before takeoff) compared against the flown path for flights in the L1 group where the contrail avoidance plan was not released (``L1 not L2'' flights). The minimum cost counterfactual plan matches better to the flown values.}
\label{fig:counterfactual_plans}
\end{figure}

In this section, we perform a counterfactual analysis to confirm that the reduction in observed contrail formation between the L2 and L3 groups is due to contrail avoidance. To perform this counterfactual analysis, we used the non-avoidance plans as a counterfactual for comparison. \autoref{fig:counterfactual_plans} shows the results of the analysis described in \autoref{sec:counterfactual_methods} to select which non-avoidance plan to use as a counterfactual. Since the ``minimum cost'' plan compares best against the flown values, we use this plan for subsequent analysis.

\begin{figure}
 \centering
        \includegraphics[width=0.7\textwidth]{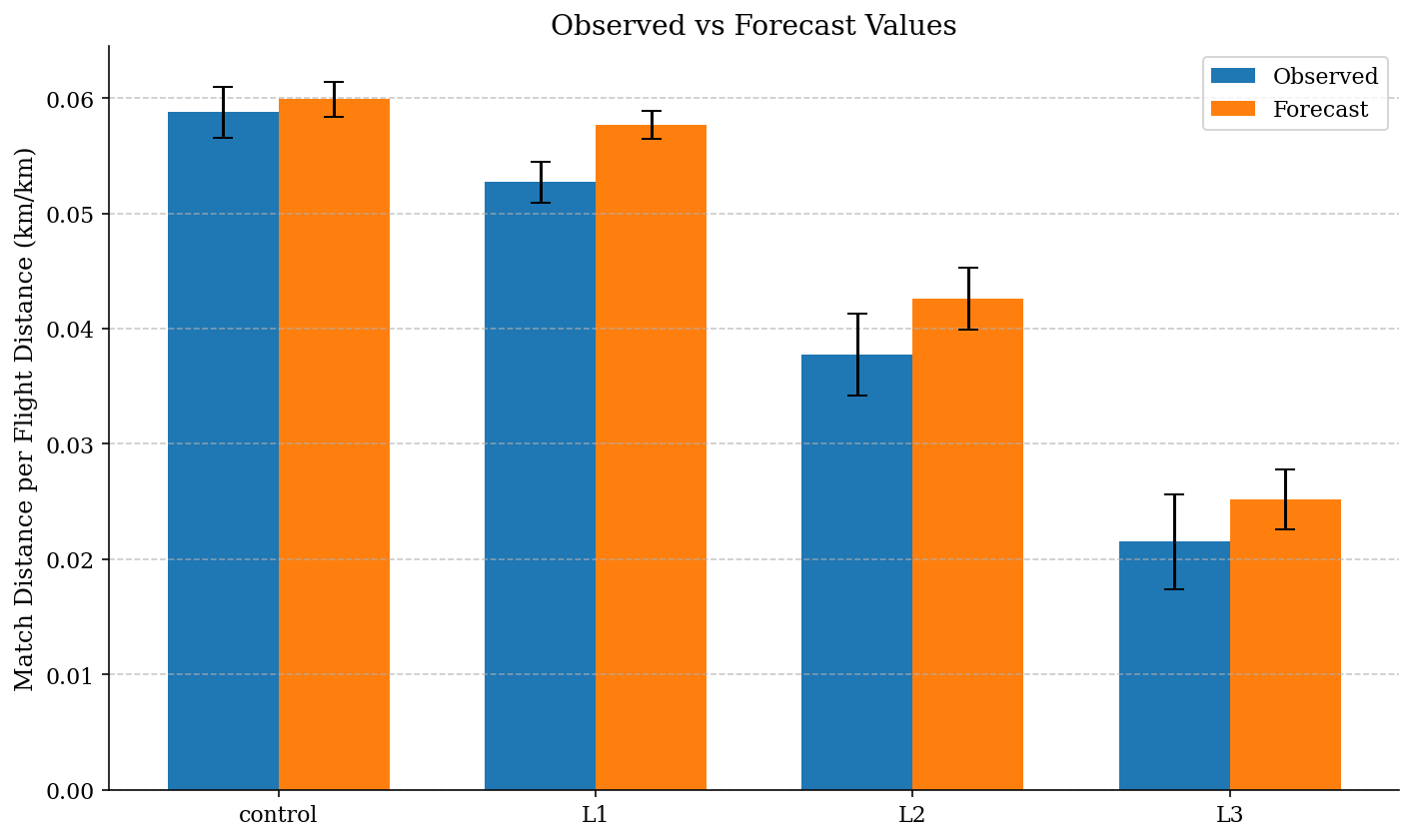}
 \caption{Comparison of the ML formation probability forecast vs. observed match distance per flight distance for the control and 3 treatment groups in the trial. The value of the bar is the mean match distance per flight distance for the group, with error bars showing the standard deviation. For each group, the forecast is well calibrated to the observations.}
\label{fig:forecast_calibration}
\end{figure}

\begin{table}[h]
\caption{Contrail distance reduction factors of the two per-protocol groups evaluated against the counterfactual plans. Expected contrail distance uses the counterfactual plan. Observed results are reproduced here from \autoref{contrail_table} as a baseline. The values in brackets represent the 95\% confidence interval of the column quantity.}
\centering
\renewcommand{\arraystretch}{2}
\begin{tabularx}{\textwidth}{l Y Y}
\toprule
 & \textbf{L2} & \makecell[c]{\textbf{L3}} \\ \hline
 
\makecell[c]{\textbf{Observed} \\ Treatment vs. \textit{Control} (\textbf{baseline})} & 
\makecell{29.5\% \\ \scriptsize {[9.7\%, 46.0\%]}} &
\makecell{62.0\% \\ \scriptsize {[47.1\%, 75.3\%]}} \\

\makecell[c]{\textbf{Expected Contrail Distance} \\ Treatment vs. \textit{Counterfactual}} & 
\makecell{$\mathbf{25.0\%}$ \\ \scriptsize {$[11.5\%, 36.8\%]$}} & 
\makecell{$\mathbf{49.2\%}$ \\ \scriptsize {$[34.9\%, 60.9\%]$}} \\



\end{tabularx}
\label{counterfactual_table}
\end{table}

Using this counterfactual, we once again evaluated the contrail results in the L2 and L3 group. Since the counterfactual plan was not flown, we cannot use observations for this analysis. Instead, we evaluate the expected contrail distance from the contrail forecast. As discussed in Sec.~\ref{sec:forecast}, the final forecast output is the product of an ML-predicted contrail formation probability and a climatological warming estimate. However, in the analysis shown in \autoref{counterfactual_table} we report on the contrail length reductions because they are more easily directly observable. The contrail probability may not in general agree with observations on an individual flight, since the probability ranges from 0-1 while the observation-derived label from the CoAtSaC algorithm is either 0 (no contrail observed) or 1 (contrail observed). However, over a large number of flights these differences average out. As shown in \autoref{fig:forecast_calibration}, the ML contrail forecast probability is well calibrated to observations when aggregated over our treatment and control groups, justifying its use as a proxy for the observed results. 

The results of this analysis are shown in Table \ref{counterfactual_table}. We see that when evaluated against the counterfactual plan, the reduction factors are smaller than the reductions seen in the observed results. However, the error bars between the counterfactual reductions and the observed reductions overlap significantly, and both treatment vs. counterfactual comparisons meet the threshold for statistical significance when evaluated using the permutation test ($p<0.001$). Note that we do not use the stratified permutation test here since the counterfactual and treatment groups will have the same aircraft type distributions by construction.

 



In order to perform a similar analysis for fuel, we selected the most recently created contrail-optimized plan as the ``treatment flight plan'' and the non-avoidance plan that was created closest in time to the treatment flight plan to be the ``control flight plan''.
Similarly to the observed results, the simulated fuel comparisons do not meet the threshold for statistical significance for the L2 and L3 groups (L2 $p=0.855$, L3 $p=0.868$, permutation test).

\section{Conclusions}
This study describes a scalable large-scale Randomized Control Trial that demonstrates the feasibility and efficacy of navigational contrail avoidance within a commercial airline's standard operational workflow. By integrating contrail forecasts directly into the Flightkeys flight planning software, we enabled dispatchers to execute avoidance maneuvers without the need for bespoke communication channels. The results provide strong evidence that navigational contrail avoidance is physically effective and operationally feasible. In the intent-to-treat group (Level 1, $n=1232$), we observed a statistically significant 11.6\% $[2.1\%,20.3\%]$ reduction in observable contrail formation relative to the control group ($p=0.011$), alongside a 13.7\% $[3.1\%,23.4\%]$ reduction in climatological warming ($p = 0.006$). These figures represent the aggregate benefit across the entire network subset, including flights where no action was taken. However, the efficacy of the intervention becomes most apparent in the per-protocol analysis (Level 3, $n=112$), which isolates flights that successfully flew the contrail-optimized plans. In this group, observable contrail length was reduced ($p<0.001$) by 62.0\% $[47.1\%,75.3\%]$ and climatological warming was reduced by ($p<0.001$) 69.3\% $[56.1\%,80.9\%]$. This substantial gap between the intent-to-treat and per-protocol results highlights that while the physical mechanism of avoidance is highly effective, the operational execution --- specifically the ``take rate" of the plans --- remains a primary bottleneck for network-wide scaling.

By utilizing the CoAtSaC  automated contrail-to-flight attribution algorithm  \citep{sarna2025benchmarking}, we were able to objectively validate performance across thousands of flights using GOES-East imagery. This methodology establishes a new standard for validating the efficacy of contrail avoidance trials, proving that contrail reductions can be quantified using peer-reviewed methodologies at large scales.

In this work we are reporting a $62.0 \%$ reduction in contrails observable via geostationary satellite, but not all contrails are observable. The reductions in total contrail distance and contrail warming are likely smaller if one assumes that some unobservable contrails are less affected by the contrail avoidance performed in this test. What fraction of contrails (and contrail warming) are observable is a subject of active research. For example, \citet{euchenhofer2025contrail} compared a set of GOES ABI images to higher-resolution low-Earth orbit (LEO) satellite images and found that 42\% of the higher-resolution contrail distance was visible in the nearest GOES scan. However, we should note that the CoAtSaC algorithm \citep{sarna2025benchmarking} only requires detecting the contrail in at least two scans captured during its lifespan, so having some frames where it is detectable in LEO but not in GOES is not inherently problematic. Furthermore, since more warming contrails tend to be larger with higher optical depth, they are also generally easier to detect in a GOES-ABI image, so the fraction of warming that is observable may be larger than the fraction of contrail distance. For example, \citet{driver2024factors} used radiative transfer simulations to estimate that, in the clear sky case, 45\% of contrail distance is observable in a given GOES ABI frame, but 57\% of contrail distance and 97\% of contrail long-wave energy forcing comes from contrails observable at some point in their lifetime. Further research is needed to better quantify the fraction of contrail warming that is observable.

Operational integration presented specific challenges that influenced the trial's outcomes. The ``take rates" of 15.4\% (flights released as contrail plans vs. total candidate flights) and 7.8\% (flights released as contrail plans and flown as planned vs. total candidate flights) are relatively low. Given the complexity of even routine daily aircraft operations, there are a large number of reasons why a contrail plan might not be released, or a released plan might not be flown. In interviews with participating dispatchers, a few common cases were discussed and summarized here as follows. Dispatchers found that contrail avoidance plans sometimes involved descents and ascents in the middle of a flight, which are safe and done in coordination with air traffic control, but which the dispatchers and flight crews preferred not to do. In the North America to Europe flights included in this trial, AAL preferred to adhere to the North Atlantic Organised Track System (OTS), and Flightkeys had special user interface features to make this easy for the dispatchers. These features were for technical reasons not able to be activated for the contrail flights during this trial, making contrail avoidance more difficult near the OTS tracks, which in many cases led to dispatchers choosing not to attempt contrail avoidance on those flights. Due to such issues, the take rate of dispatchers releasing a contrail-aware flight plan ranged widely per city pair. 
Dispatchers also mentioned that their take rate was typically lower when they and the aircraft crew were busier. As this trial was executed on a voluntary basis, a dispatcher could also choose to not take part for any reason. Dispatcher priorities were established with safety as the primary focus, followed by efficiency and finally contrails. The L2/L1 take rate on a per-dispatcher basis ranged widely,
suggesting that inter-dispatcher variation is important. Since city pairs were not evenly distributed between dispatchers, this likely had an effect on the difference in take rate per city pair. Finally, dispatchers mentioned the presence of turbulence, which is a safety concern, as a reason for not considering the contrail-optimized flight plan. Further research is necessary to understand the correlation between turbulence and contrail formation.

A few software improvements could improve the take rate for future trials. In the user interface available to dispatchers for this trial, there was only a top-down view of the contrail avoidance regions, but additionally including a vertical profile view would be beneficial. This made it difficult for dispatchers to understand why certain adjustments were happening. 

Improvements to the user interface inside flight planning software could lead to a larger observed contrail reduction, even in the L3 group. Dispatchers frequently make manual adjustments to software-optimized plans, even when contrails are not considered. 
In a few (2-3) instances in the trial, dispatchers, without the benefit of the vertical profile view of the contrail avoidance regions, adjusted a contrail-optimized plan to fly through the contrail avoidance region, but these flights nonetheless ended up in the L3 group. 

Around half of the released contrail-optimized flight plans were not flown as planned, as determined by the L3 criteria. This could potentially be ameliorated by displaying contrail information in tools accessible to pilots such as the `Electronic Flight Bag' (EFB), so that pilots can also understand why maneuvers are being made. Additionally, the automated flight planning process starts roughly 24 hours prior to departure, and the dispatchers engage with the process closer to the departure time. The contrail forecast can change during that time. If a flight was initially clearing the 10 t threshold, a dispatcher could start working on a contrail plan even if later updates to the weather forecast brought the flight below the 10 t threshold. This could have also occurred in the control group and therefore should not substantially affect the results, but it might have caused flights whose non-avoidance plans had low contrail impact to end up in both the treatment and control groups. Future improvements to flight planning software could ameliorate both of these problems. Additionally, in a system-wide deployment of contrail avoidance, the efficacy of the contrail avoidance regime may improve since dispatchers would only be presented with flight plans that consider contrails, rather than being presented with multiple non-avoidance and contrail-optimized plans from which to choose.

Our analysis of fuel burn revealed a counter-intuitive result: the Level 1 treatment group appeared to consume 0.55\% less fuel than the control group (adjusted p=0.044). As detailed in Section \ref{sec:costs}, this finding is likely driven by residual confounding within the randomization structure rather than a genuine fuel benefit of contrail avoidance. This ``negative cost" result, along with the $\approx 2$ percentage-point-wide confidence intervals for fuel reduction in the L2/L3 groups, underscores the difficulty of isolating the fuel penalty of avoidance (estimated at 0.1-0.2\% across all of an airline's flights \citep{teoh2020mitigating,frias2024feasibility,dean2025impact}) against the high variance of real-world fuel consumption data. Larger sample sizes and stricter randomization protocols in future trials will be necessary to precisely quantify the fuel penalty trade-off.

Finally, while our counterfactual analysis suggests that the observed reductions were not driven by dispatchers selecting flights that would have formed fewer contrails even without our intervention, the infrequent randomization of city pairs remains a limitation. To further establish causality and refine the cost-benefit analysis, future work could expand beyond the North Atlantic to a global scope, increase the frequency of randomization, and further automate the optimization workflow to reduce labor burden on dispatchers. Future work should also explore different operational concepts for contrail avoidance such as coordination between airlines and Air Navigation Service Providers (ANSPs), ANSP-led trials, or coordination between multiple airlines. Despite these limitations, this trial provides a statistically robust point of evidence that commercial aviation can successfully avoid warming contrails using existing operational infrastructure.

\begin{appendices}

\section{Additional Observed Contrail Values} \label{appendix:observed_values}
\begin{table}[ht]
\centering
\small
\renewcommand{\arraystretch}{1.4}
\caption{Observed contrail formation results and total observed distances.}
\label{tab:contrail_distance}

\begin{tabular}{l c c}
\toprule
\textbf{Group} & \textbf{$N$} & \textbf{Total Observed Contrail Distance (km)} \\
\midrule
Control         & 1172 & 299,317 \\
\addlinespace
Treatment (L1)  & 1232 & 281,214 \\
\addlinespace
Treatment (L2)  & 190  & 32,223 \\
\addlinespace
Treatment (L3)  & 112  & 10,640 \\
\bottomrule
\end{tabular}
\end{table}

Table~\ref{tab:contrail_distance} shows additional observed contrail values.

\section{Modifications to the Contrail-to-Flight Attribution Algorithm} \label{appendix:altitude}
For the validation system for the trial described in this paper, we used a modified version of the contrail-to-flight attribution algorithm described in \citet{sarna2025benchmarking}. The modified version of this algorithm is also used for generating training data for the ML contrail formation forecast.

\subsection{Contrail Altitude Model}
We developed a system to infer the altitude of a contrail in GOES imagery, similar to the method introduced in \citet{meijer2024contrail}. The system uses a convolutional neural network (CNN) that takes radiances from tiles of the GOES imagery as input and infers an altitude at each pixel in the tile. This CNN is trained using altitudes from flight-to-contrail attributions as labels. Specifically, during the attribution process, the advection of exhaust plumes from flights are simulated in 3D space from their original locations to the times of satellite images, as described in \citep{sarna2025benchmarking}. If the attribution algorithm finds a matched flight for a given detected contrail, the advected altitude of that flight is used as the label for the pixels spanned by the detected contrail. Detected contrails that do not have a matched flight are not included in the training data. The loss is only applied on pixels where the dataset includes a detected contrail with a corresponding matched flight.

\begin{figure}
 \centering
        \includegraphics[width=1\textwidth]{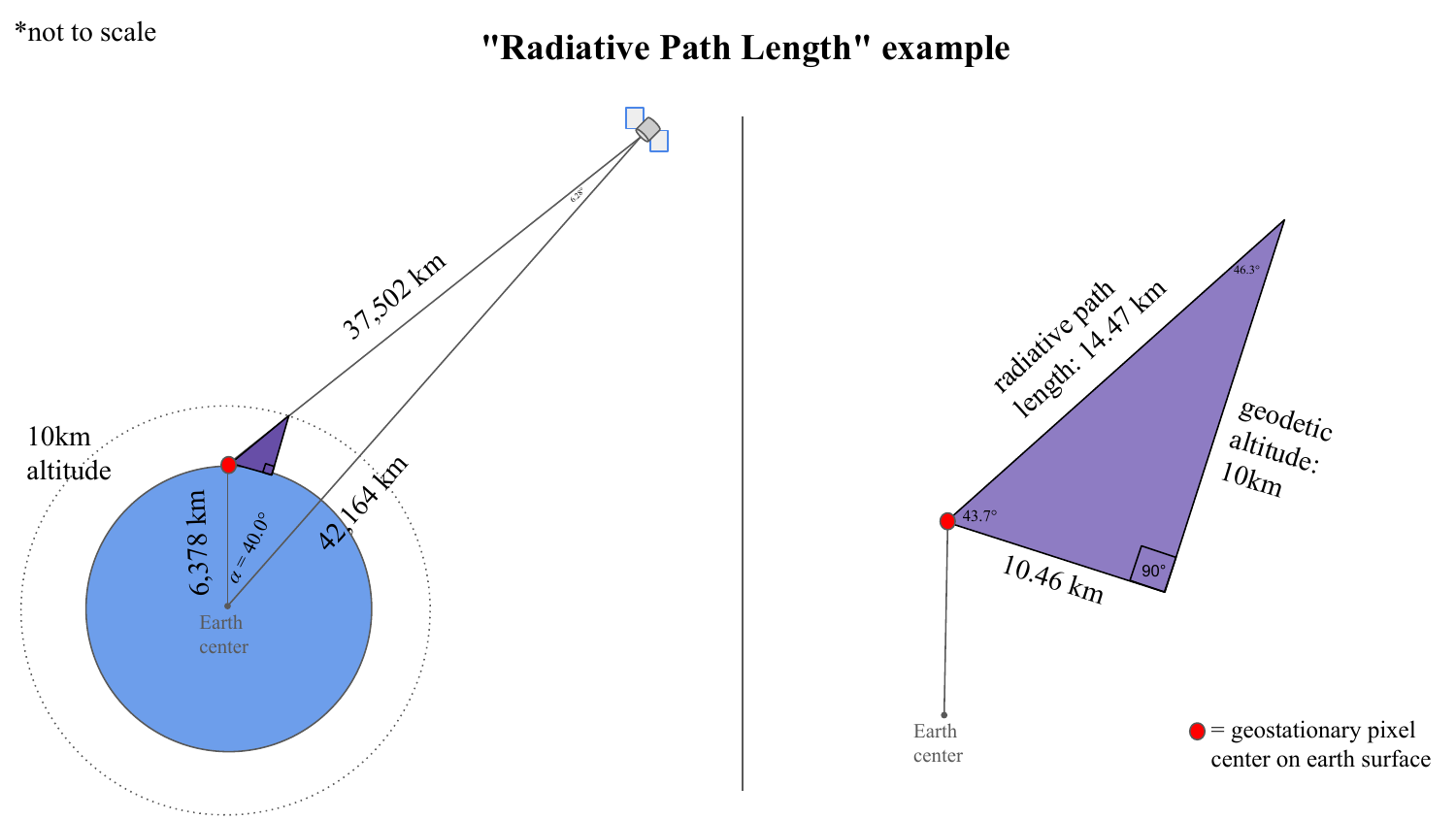}
 \caption{Example of a transformation from geodetic altitude to radiative path length. This simplified example considers a contrail that is over the Equator at 40$^{\circ}$ longitude away from the nadir of the geostationary satellite. Its geodetic altitude is 10 km, which is translated into a radiative path length of 14.47 km using the triangle formed by the satellite, contrail, and center of the Earth and the purple triangle formed by the contrail, the point directly underneath it on the surface, and the point on the surface that the satellite is viewing for the pixel in which the contrail appears. Side lengths and angles other than the 40 degrees longitude, Earth radius, and geostationary orbit distance are derived via trigonometric identities.}
\label{fig:radiative_path_length}
\end{figure}

In order to account for the effects of viewing angle on the perceived altitude by the satellite, we transform the advected barometric altitude of the flight into a ``radiative path length,'' which allows the altitude estimation model to take into account the amount of atmosphere through which the satellite-sensed radiation traveled between the earth surface and the observed contrail. To perform this transformation, we first interpolate the advected pressure level in ERA5 reanalysis data to find its corresponding geopotential height, and then convert that to geodetic height by dividing it by the acceleration due to Earth's gravity. We then apply trigonometric identities (visualized for a simplified 2D case in Figure~\ref{fig:radiative_path_length}) to translate the geodetic height into a radiative path length. After the model performs inference, the inverse of this trigonometric transformation is used to translate the radiative path length back to a geodetic altitude.

During the training process, the model is validated against a held-out set of flight-to-contrail attributions. This dataset has the advantage of wide spatiotemporal coverage, but has the disadvantage of incorporating the noise introduced by the inaccuracies of the attribution algorithm. In order to ensure that the model performs well and generalizes, we additionally evaluate its performance on the LiDAR colocation dataset from \citet{meijer2024contrail} that provides high-accuracy altitude estimations from the CALIOP instrument onboard the CALIPSO satellite. On this dataset, our model achieves a root mean squared error (RMSE) of 789$m$, compared to the RMSE of 570$m$ from \citet{meijer2024contrail}. We chose not to use the CALIOP data for training data generation in the model since CALIPSO was decommissioned in 2023.

At inference time, the model produces an inferred altitude for every pixel in the input satellite image, regardless of whether there is a contrail present. In order to use these inference results in the downstream attribution algorithm, we join the inferred altitudes with detected contrails from \citet{ng2023contrail}. Each linearized contrail is assigned the mean altitude of the inference results from the pixels it contains.

\subsection{Modified Contrail-to-Flight Attribution Algorithm}
At a high level, the system described in \citet{sarna2025benchmarking} begins by considering attributions in each frame of the satellite imagery and looks for patterns in how those single-frame attributions evolve over time through multiple frames of the satellite image. The key modification we make to this algorithm is to filter out single-frame attributions where the advected altitude of the flight is significantly different from the inferred altitude of the detected contrail as determined by the CNN. We set the threshold as 

$$2 * \sqrt{(altitude\;estimation\;RMSE)^2 + (vertical\;advection\;RMSE)^2},$$

\noindent which results in $2 * \sqrt{800m^2 + 111m^2} = 1615$ m. The vertical advection RMSE is computed independently of the advection age by comparing the advected altitude to the known synthetic altitude of contrails in SynthOpenContrails \citep{sarna2025benchmarking}. Therefore, if in a single satellite frame the inferred contrail altitude differs from the advected flight altitude by more than $1615$ m, the single-frame attribution is not considered as a candidate match for the subsequent multitemporal aspects of the algorithm. In order to evaluate the effect of this modification on the performance on CoAtSaC, we designed a version of the evaluation of CoAtSaC against SynthOpenContrails that assumes an RMSE of the altitude estimator and assigns the synthetic predicted altitude of a contrail to a value sampled from a Gaussian where the mean is the known altitude of the contrail and the standard deviation is the RMSE of the altitude estimator. This synthetic predicted altitude is used in place of the inferred altitude of the detected contrail in the modified algorithm described above. With this change, the ``flight precision'' on the tracking subset of SynthOpenContrails improves from 71.4\% to 79.4\%.

\section{City Pair Assignments}
Table \ref{tab:route_assignments} shows the city pair assignments by week in the treatment and control groups.

\begin{table}[htbp]
    \centering
    \caption{Flight route assignments by week and group. Week 1 began on Wednesday, January 15, 2025. Each week of the trial started on Wednesday and ended on the following Tuesday.}
    \label{tab:route_assignments}
    \begin{tabularx}{\textwidth}{l >{\raggedright\arraybackslash}X >{\raggedright\arraybackslash}X}
        \toprule
        \textbf{Week} & \textbf{Treatment Group Routes} & \textbf{Control Group Routes} \\
        \midrule
        \makecell[l]{Week 1\\(Pretrial,\\Not Considered\\in Analysis)} & JFK-LHR & --- \\
        \addlinespace
        Week 2 & JFK-LHR, MIA-MAD, MIA-BCN, PHL-MAD & JFK-MAD, CLT-FRA, JFK-MXP, DFW-CDG \\
        \addlinespace
        Week 3 & JFK-LHR, MIA-MAD, MIA-BCN, PHL-MAD, PHL-BCN, CLT-MUC, CLT-FRA, MIA-LHR & ORD-LHR, CLT-LHR, CLT-MAD, PHL-LIS, PHL-CDG, PHL-AMS, JFK-MAD, DFW-CDG \\
        \addlinespace
        Week 4 & JFK-LHR, MIA-MAD, MIA-BCN, PHL-MAD, PHL-BCN, CLT-MUC, CLT-FRA, MIA-LHR, DFW-LHR, ORD-LHR, JFK-CDG, MIA-CDG, DFW-CDG, PHL-CDG, PHL-ZRH, JFK-MXP & PHL-LHR, PHL-LIS, CLT-MAD, BOS-LHR, DFW-FRA, LAX-LHR, CLT-LHR, PHL-AMS, PHL-DUB, JFK-MAD \\
        \addlinespace
        Week 5--7 & CLT-MUC, DFW-FRA, JFK-CDG, JFK-MAD, JFK-MXP, MIA-LHR, MIA-MAD, ORD-LHR, PHL-AMS, PHL-BCN, PHL-LHR, PHL-LIS, PHL-MAD, PHL-ZRH & JFK-LHR, MIA-BCN, CLT-FRA, MIA-CDG, DFW-CDG, PHL-CDG, CLT-MAD, BOS-LHR, LAX-LHR, CLT-LHR, PHL-DUB \\
        \addlinespace
        Week 8--10 & CLT-FRA, CLT-MAD, CLT-MUC, DFW-CDG, JFK-MAD, JFK-MXP, MIA-CDG, MIA-LHR, MIA-MAD, ORD-LHR, PHL-AMS, PHL-BCN, PHL-DUB, PHL-ZRH & JFK-LHR, MIA-BCN, PHL-MAD, JFK-CDG, PHL-CDG, PHL-LIS, PHL-LHR, BOS-LHR, DFW-FRA, LAX-LHR, CLT-LHR \\
        \addlinespace
        Week 11--12 & MIA-LHR, PHL-DUB, BOS-LHR, MIA-MAD, PHL-MAD, PHL-CDG, DFW-FRA, JFK-CDG, JFK-MAD, PHL-LIS, PHL-ZRH, DFW-CDG & JFK-LHR, MIA-BCN, PHL-BCN, CLT-MUC, CLT-FRA, MIA-CDG, JFK-MXP, PHL-LHR, CLT-MAD, PHL-AMS, DFW-FCO \\
        \addlinespace
        Week 13--14 & JFK-MXP, PHL-AMS, PHL-MAD, PHL-DUB, BOS-LHR, PHL-ZRH, MIA-LHR, MIA-MAD, PHL-BCN, DFW-MAD, PHL-LHR, MIA-BCN, PHL-FCO, DFW-FCO & JFK-LHR, CLT-MUC, CLT-FRA, JFK-CDG, MIA-CDG, DFW-CDG, PHL-CDG, PHL-LIS, CLT-MAD, DFW-FRA, JFK-MAD, PHL-ATH, DFW-BCN, JFK-BCN, JFK-FCO \\
        \addlinespace
        Week 15--17 & PHL-ZRH, PHL-LIS, CLT-FRA, DFW-BCN, CLT-MAD, DFW-CDG, PHL-DUB, PHL-LHR, PHL-BCN, JFK-MXP, PHL-FCO, PHL-MAD, PHL-CDG, JFK-BCN & JFK-LHR, MIA-BCN, CLT-MUC, MIA-LHR, JFK-CDG, MIA-CDG, BOS-LHR, DFW-FRA, PHL-AMS, JFK-MAD, DFW-FCO, PHL-ATH, DFW-MAD, JFK-FCO \\
        \bottomrule
    \end{tabularx}
\end{table}

\section{Flight Adherence Visualizations}  \label{appendix:viz_examples}
\begin{figure}[htbp]
     \centering
     \includegraphics[width=\textwidth]{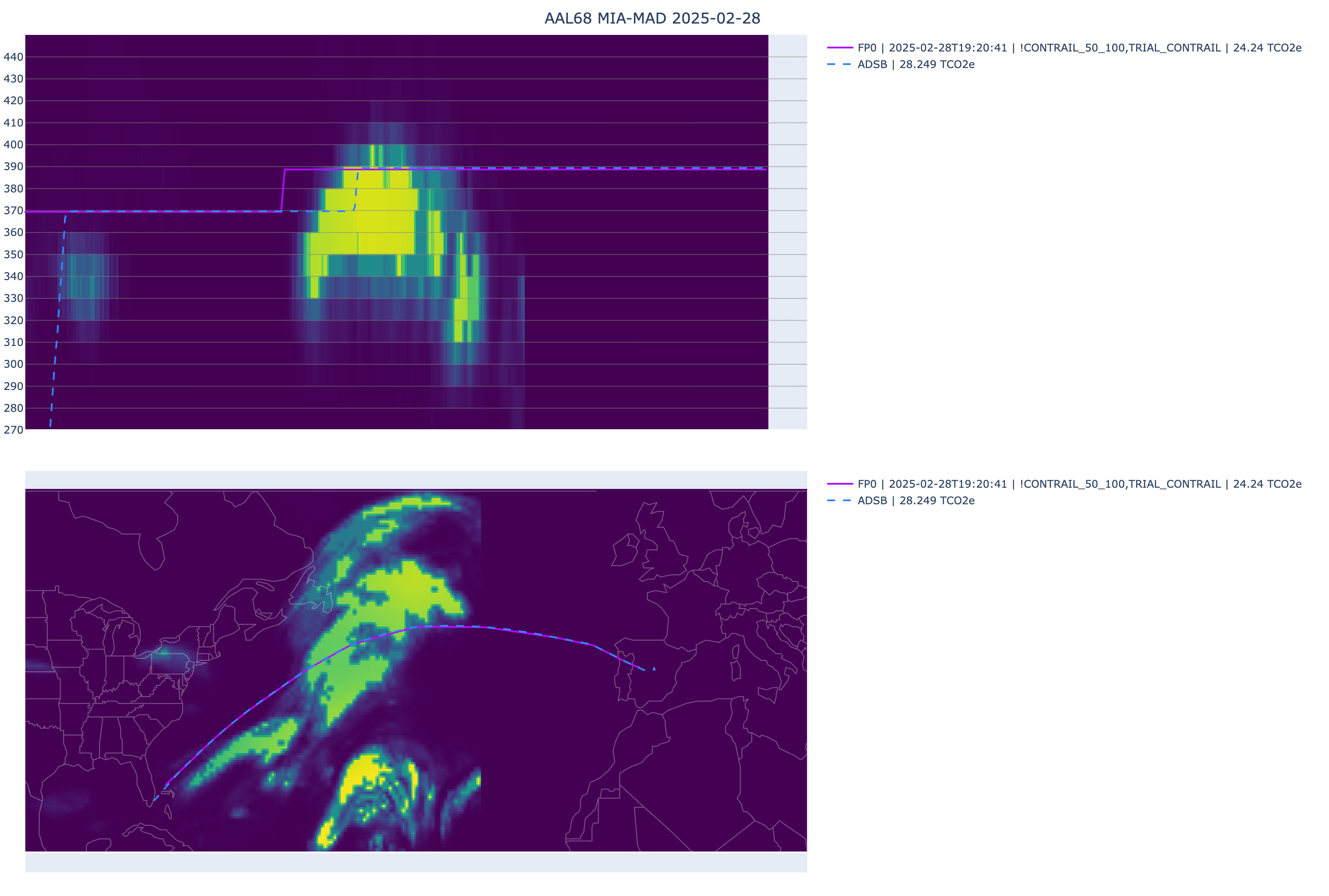}
     \caption{Example of an L2 flight that is not included in the L3 set. Dashed blue lines represent the flown path, and solid purple lines represent the planned path. The top image shows the vertical altitude profile, where the green colors represent the contrail forecast interpolated to the flown path. The bottom image shows the lateral profile, with the contrail forecast at a fixed 35,000 ft.}
     \label{fig:l2_viz}
\end{figure}

\begin{figure}[htbp]
     \centering
     \includegraphics[width=\textwidth]{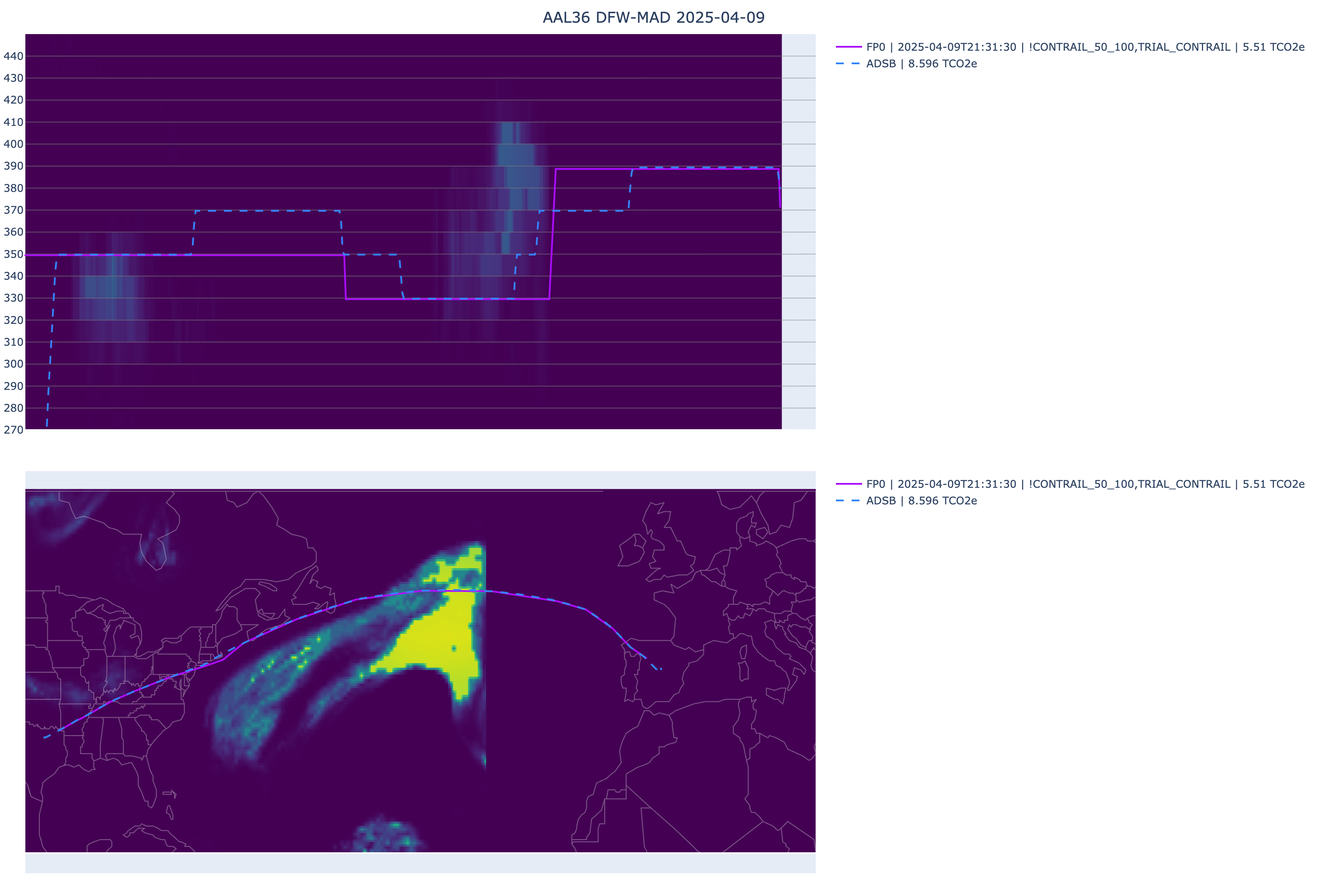}
     \caption{Example of an L3 flight. Despite imperfect adherence to the vertical profile, the difference in forecasted contrail impact is not large enough to fail the L3 inclusion check. Dashed blue lines represent the flown path, and solid purple lines represent the planned path. The top image shows the vertical altitude profile, where the green colors represent the contrail forecast interpolated to the flown path. The bottom image shows the lateral profile, with the contrail forecast at a fixed 35,000 ft.}
     \label{fig:l3_viz}
\end{figure}

\autoref{fig:l2_viz} and \autoref{fig:l3_viz} show examples of the visualization used by the authors of this work to assess flight plan adherence.

\section{Aircraft Type Distributions}
\begin{figure}
 \centering
        \includegraphics[width=0.7\textwidth]{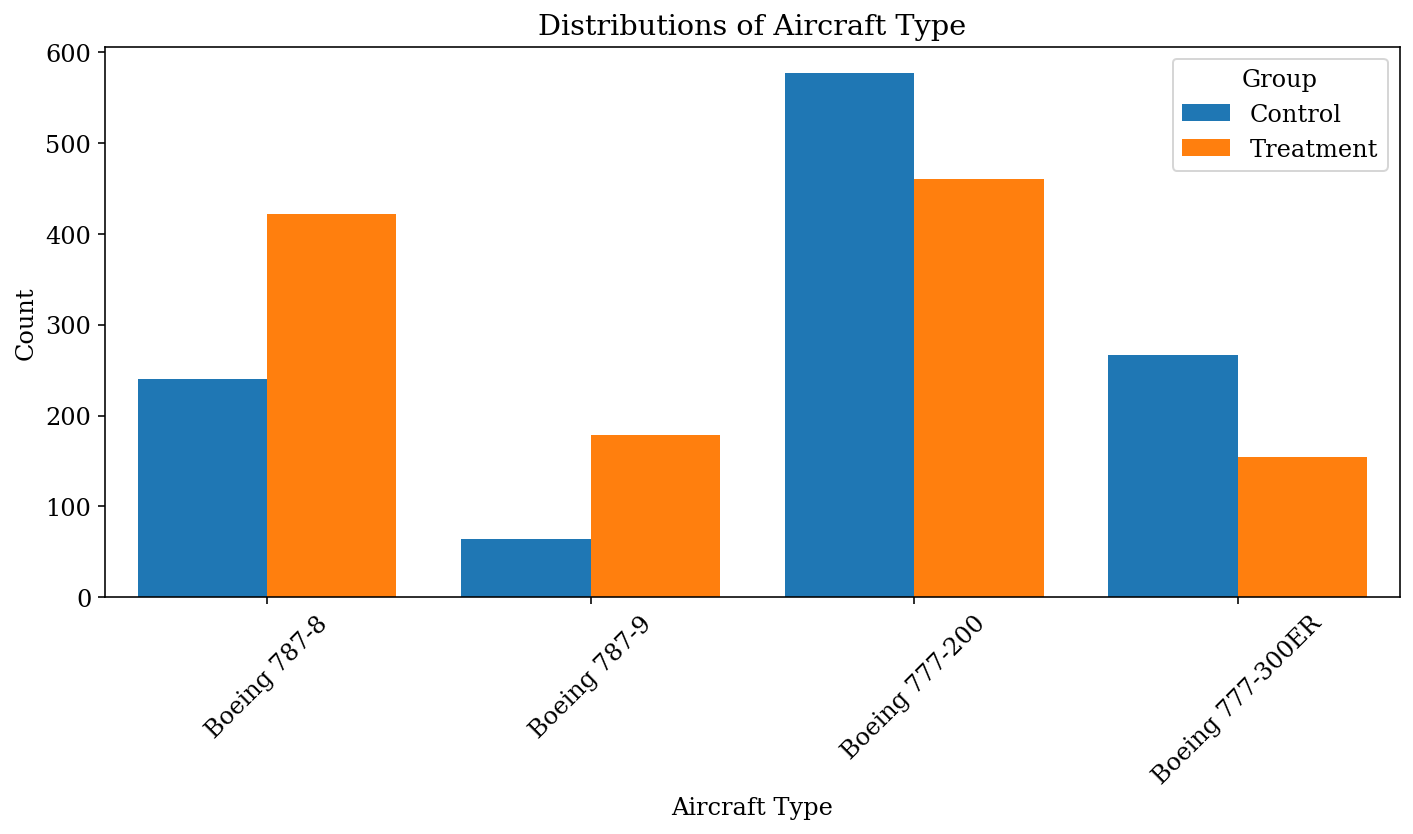}
 \caption{Distributions of aircraft types across the treatment and control groups before the adjustment described in \autoref{sec:adjustment}.}
\label{fig:aircraft_type_distribution}
\end{figure}

Fig.~\ref{fig:aircraft_type_distribution} shows the distributions of aircraft type in the unadjusted treatment and control groups.

\end{appendices}

\section{Author contributions}
The study scope and randomized trial design were developed by T.S., T.D., S.G, and A.S.W. in collaboration with J.B., M.G., R.G., and A.P. J.B., M.G., R.G., and A.P. led the operational implementation of the study at the airline, including communications with participants in the trial. A.M.F. and R.Z. developed the integration of the contrail forecast into the flight planning software. A.S., T.S., and S.G. compiled and pre-processed the necessary flight and satellite data for the analysis. T.S. and A.S.W. conducted the statistical analysis. T.A., P.H., K.M., T.R., D.S., and M.S. provided scientific advising throughout the development of the study and the compiling of the results. T.S. interpreted the results and wrote the manuscript, including significant contributions from T.D., T.R., and S.G. All authors reviewed the manuscript.

\section{Competing interests}
The authors declare the following competing interests: As denoted by their affiliations, some authors are employed by Google LLC, Flightkeys GmbH, and American Airlines. Contrails.org is operated by Breakthrough Energy, a family of organizations and activities committed to transitioning the world to net zero by 2050. All other authors declare no competing interests.

\section{Acknowledgements}
We dedicate this paper to the memory of Darcy Freeman, a dispatcher who released many contrail flights during the course of this trial. We thank The International Council on Clean Transportation (ICCT) for providing grant funding to AAL to catalyze the trial. We would also like to thank Greg Thorstensen, Daniel Royal III, and Derek Voege, dispatchers who released many contrail flights and provided us with valuable insights about the opportunities and challenges of contrail avoidance; John Dudley, Deborah Hecker, Rondeau Flynn, and Kevin Macelhaney, pilots who supported the development and execution of our trial; Kevin Johle, who provided critical technical support for the operation of the trial; and Ethan Klapper, Chris D'Aloia, Ben Halle, and Thomas Adler for reviewing the manuscript. The authors of this work used Google Gemini AI to draft parts of the manuscript, which were then reviewed and heavily edited by the authors.

\bibliographystyle{jphysicsB}
\bibliography{sample}

\end{document}